# A PERFORMANCE ANALYSIS FOR UMTS PACKET SWITCHED NETWORK BASED ON MULTIVARIATE KPIS

Ye Ouyang and M. Hosein Fallah, Ph.D., P.E.
Howe School of Technology Management
Stevens Institute of Technology, Hoboken, NJ, USA 07030
youyang@stevens.edu
hfallah@stevens.edu

## ABSTRACT

*Mobile data services are penetrating mobile markets rapidly. The mobile industry relies heavily on data service to replace the traditional voice services with the evolution of the wireless technology and market. A reliable packet service network is critical to the mobile operators to maintain their core competence in data service market. Furthermore, mobile operators need to develop effective operational models to manage the varying mix of voice, data and video traffic on a single network. Application of statistical models could prove to be an effective approach. This paper first introduces the architecture of Universal Mobile Telecommunications System (UMTS) packet switched (PS) network and then applies multivariate statistical analysis to Key Performance Indicators (KPI) monitored from network entities in UMTS PS network to guide the long term capacity planning for the network. The approach proposed in this paper could be helpful to mobile operators in operating and maintaining their 3G packet switched networks for the long run.*

## KEYWORDS

*UMTS; Packet Switch; Multidimensional Scaling; Network Operations, Correspondence Analysis; GGSN; SGSN; Correlation Analysis.*

## 1. INTRODUCTION

Packet switched domain of 3G UMTS network serves all data related services for the mobile subscribers. Nowadays people have a certain expectation for their experience of mobile data services that the mobile wireless environment has not fully met since the speed at which they can access their packet switching services has been limited. Mobile operators realize that if they are to succeed in today's wireless communications landscape, they must address the quality of service for their packet service users. Simply adding more bandwidth to accommodate increased packet switching traffic is an expensive alternative. Hence, the mobile operators are faced with the issue of how to do more with less? The initial answer is to ensure the network is operating optimally before one considers further capital investment in expanding the network infrastructure.

For a network administrator, the traditional network operation and maintenance (O&M) pattern follows a cycle: If a problem is encountered, from hardware or software failures to network congestions, the technician issues a ticket, debugs the network, and fix the problem and operation continues. This mode of operation may be adequate for ensuring timely and quality service of data traffic in a short run. However it does not help mobile operators effectively and actively forecast and prevent potential problems in packet switched network in advance. This paper offers an approach to help mobile operators shape O&M policies for the long run via applying multivariate statistical modeling to the KPIs obtained from UMTS packet switched network.

## 2. LITERATURE REVIEW

The current literature provides many practical tools and theoretical methods to design, plan and

International Journal of Next-Generation Networks (IJNGN), Vol.2, No.1, March 2010

dimension UMTS PS network. Also no previous literature provides a unified approach to estimate the performance for UMTS packet switched network from the aspect of multivariate analysis. The specifications of 3GPP from [2], [3], [4] and [5] define the architecture, topology, and services of UMTS Packet Switched (PS) network. Figure 1 displays the key network entities and critical links in the UMTS core network. The specifications such as [6], [7], [8], [9] and [10] define the protocol stacks of the interfaces in PS domain and the tunneling encapsulation process between Serving GPRS Support Node (SGSN) and Gateway GPRS Support Node (GGSN). The first ten references are the technical guidelines to construct the trial PS network. Reference [18], [19], and [20] propose some optimized network architecture of UMTS core networks and provide a unified approach to calculate the throughput or traffic for UMTS packet switched network

In addition, [11], [12], [13], and [16] introduce the possible methods of multivariate analysis, some of which are adopted in the article to estimate the performance of UMTS PS network and forecast the potential failures exited in the network. Literatures [14], [15], and [17] are focused on those economic and policy issues in US telecommunication industry via empirical studies in which the various methods of multivariate analysis applied.

## 3. PACKET SWITCHED DOMAIN IN 3G UMTS

Packet Switched (PS) domain and Circuit Switched Domain compose the Core Network (CN) of a 2G Global Systems for Mobile Communications (GSM) or a 3G UMTS network. Whether in 2G or 3G phase, the CN plays an essential role in the mobile network system to provide such important capabilities as mobility management, call and session control, switching and routing, charging and billing, and security protection.

In R99 version, the first version of 3G UMTS network, the CN domain still consists of the same network entities (NE) and the same network architecture as that in GSM phase. However, there is a change in the circuit switched domain of R4, the second version of UMTS, which supports a networking mode where bearer is separated from control. Meanwhile multiple bearer modes such as ATM/IP/TDM are supported by CN. Consequently the Mobile Switching Center (MSC) in GSM/UMTS R99 is split into two NEs: MSC Server (MSS) and Media Gateway (MGW). We should note that no changes happen in packet switched domain from R99 to R4 except for a new Iu-PS interface which is used to connect PS domain with 3G radio access network (RAN).

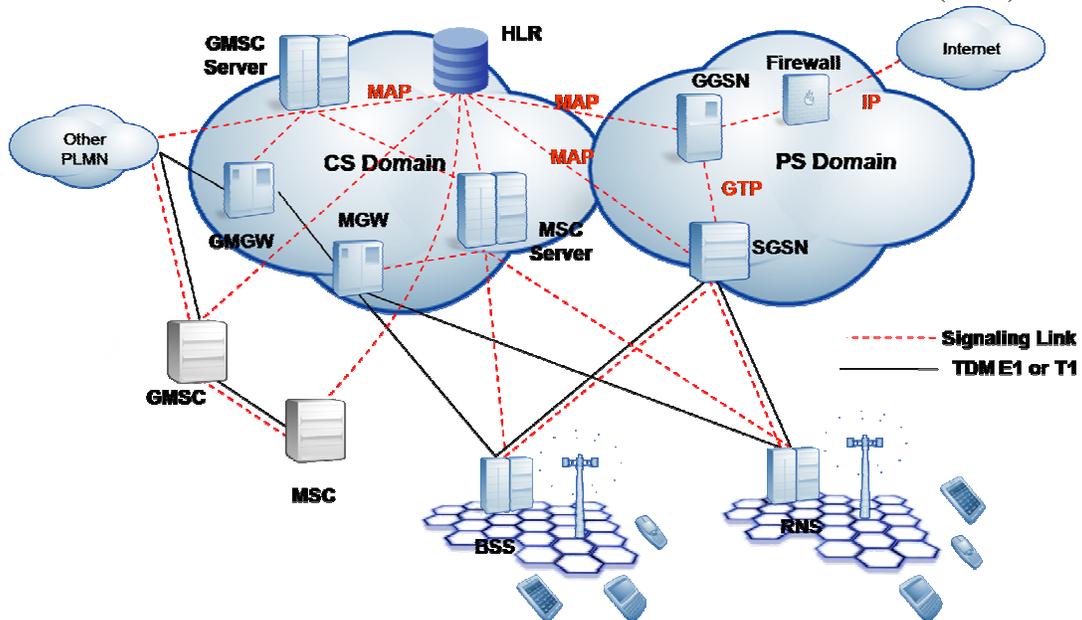

Figure 1. Topology of UMTS CN: CS+PS domain

The CN in UMTS is logically classified into the circuit switched domain (CS) and packet switched domain (PS). The CS domain includes such logical NEs as MSC Server, MGW, Visitor Location Register (VLR) integrated in MSC Server physically, Home Location Register (HLR), Authentication Center (AUC), and Equipment Identity Register (EIR). The packet switched domain (PS) includes Serving GPRS Support Node (SGSN) and Gateway GPRS Support Node (GGSN). More specifically, PS domain consists of data service NEs: SGSN and GGSN as well as auxiliary NEs like Charging Gateway (CG), Border Gateway (BG) and Domain Name System Server (DNS), and different service platforms attached to PS domain. Figure 1 displays the topology of UMTS CN with the logical NEs mentioned above.

Packet Switched domain physically consists of SGSN, GGSN, and Charging Gateway. Below is a short description of these NEs.

SGSN is responsible for the delivery of data packets from and to MSs within its serving area. Its tasks include packet routing and transfer, mobility management (attach/detach and location management), logical link management, and authentication and charging functions. Its interfaces include Iu-Ps interface connecting to RNC, Gn/Gp interface to GGSN, Gr interface to HLR, Gs interface to MSC Server or MSC, Gd interface to Short Message Center (SMC), and Ga interface to Charging Gateway.

GGSN is a gateway between UMTS PS/GPRS network and external data networks (e.g. Internet). It performs such functions as routing and data encapsulation between a MS and external data network, security control, network access control and network management. From UMTS PS/GPRS aspect, a MS selects a GGSN as its routing device between itself and external network in the activation process of PDP context in which Access Point Name (APN) defines the access point to destination data network. From external data network aspect, GGSN is a router that can address all MS IPs in UMTS PS/GPRS network. GGSN provides Gc interface to connect with HLR, Gn/Gp interface with SGSN, Gi interface with external data networks, and Ga interface with CG.

Charging Gateway is the billing unit for PS domain. Sometimes coupled together with SGSN, it collects, merges, filters and stores the original Call Detail Record (CDR) from SGSN and communicates with billing center, and then transfers sorted CDR to billing center.

## 4. SERVICE MONITORING MODEL FOR UMTS PS NETWORK

UMTS Packet Switched (PS) network is a typical data network in which data traffic, particularly with streaming media services, is live, extremely time sensitive to delay, latency and jitter, non-tolerant of congestion. For example, a small minority of packet service subscribers running FTP, streaming video or peer-to-peer (P2P) file sharing applications can generate enough traffic to congest UMTS PS networks and impact the majority of subscribers using interactive Web browsing and E-mail applications.

In the past network operation and maintenance was focused more on monitoring the entire throughput. The UMTS PS model for service monitoring shall be capable of monitoring and capturing the necessary KPI data at the service level in addition to the network level. In the model, various types of service packet enter PS core domain via Iu-PS interface, the entry port of SGSN. After the encapsulated tunneling transport between SGSN and GGSN, the packets are delivered out to external network via the exit: Gi interface in GGSN. Hence the data monitoring starts from interface Iu-PS, the entry port of SGSN, and ends in interface Gi which is the exit of GGSN. The monitored KPIs for the model include two types of parameters: QoS/performance parameters and service parameters, the former of which includes delay, jitter, packet loss, throughput, and utilization; while the latter includes the throughput of all types of services going through SGSN and GGSN. Figure 2 below depicts the service model of UMTS PS network for performance monitoring. Different from traditional instant network monitoring, the UMTS PS model for service monitoring shall achieve:

International Journal of Next-Generation Networks (IJNGN), Vol.2, No.1, March 2010- A long run view of the PS service the user is experiencing;
- Service-level quality and performance metrics which are affected by the traffic as well as vendors equipment (SGSN and GGSN);
- Correlation of fault and performance data captured over a long period to identify the potential service affecting outages;
- Consolidated utilization and performance data that can be applied for future network expansion planning.

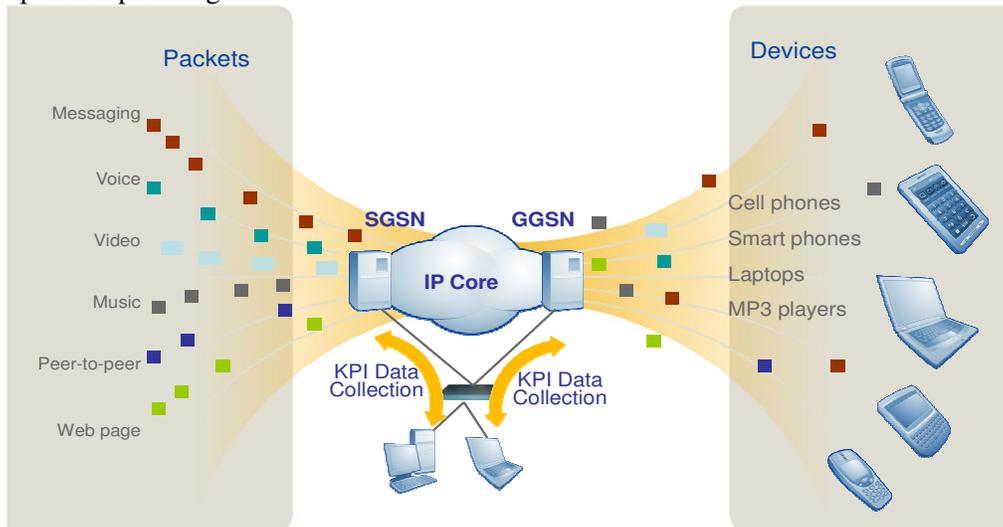

Figure 2. UMTS PS Model of Service Monitoring

## 5. MULTIVARIATE STATISTICAL METHODS APPLIED TO KPIS

The following methods can be applied to the collected UMTS PS performance (QoS) parameters and service parameters. Below is a short introduction to the methods we will be using. For the detail algorithms please see the Appendix A.

*A. Correlation*

Correlation measures the strength and direction of the relationship between two or more variables. Correlation is a standardized measurement that generates an easily interpretable value (Correlation Coefficients: r) ranging from -1.00 to +1.00. Correlation as applied in the next section will analyze the relationship between QoS performance parameters and service parameters. The correlation results may identify the service-level quality and performance metrics and reveal the impact of different service types on the UMTS PS network performance.

*B. Factor Analysis*

The core purpose of factor analysis is to describe the covariance relationships among many variables in terms of a few underlying, but unobservable, random quantities called factors. F actor analysis identifies the key variables and detects the structure in the relationships between variables, that is, to classify variables. Factor analysis can be considered an extension of principle component analysis, both attempting to approximate the covariance matrix represented by Σ. However, the approximation based on the factor analysis model is more elaborate.

*C. Multidimensional Scaling*

Multidimensional scaling displays multivariate data in low-dimensional space. Its primary target is to fit the original data into a low dimensional coordinate system such that any distortion caused by a reduction in dimensionality is minimized. Multidimensional scaling in our case deal with the following problem: For a set of observed similarities or differences between pairs of observations, we can find a representation of the samples in two dimensions such that their proximities in the new space nearly match the original similarities.

*D. Correspondence Analysis*

83

Correspondence analysis is a graphical procedure for representing associations in a table of frequencies or counts. To illustrate this assume $X = \{x_{ij}\}_{m \times n}$ is the matrix of m observable values of network parameters in n samples. Based on the factor analysis, one can obtain the eigen values for the m parameters. Then if we select p common factors ($p \leq m$) to calculate their corresponding eigen vectors {$eig_1$, $eig_2$,…$eig_p$}, the loading matrix last obtained is the common factor scores for the m parameters. Similarly the common factor scores for sample time spots can also be obtained. The last step is to reflect the factor scores for UMTS PS network parameters and sample times into a two dimension coordinates Using this approach we are able to investigate the UMTS PS network QoS in different time intervals to develop operation policies to optimize the performance of the current network.

*E. Cluster Analysis*

Cluster analysis is an exploratory data analysis tool which aims at sorting different objects into groups in a way that the degree of association between two objects is maximal if they belong to the same group and minimal otherwise. Cluster analysis simply discovers structures in data without explaining why they exist. Cluster analysis can be used to discover structures in observed in KPI samples without providing an explanation/interpretation. Cluster Analysis and Multidimensional Scaling used together can help to investigate the similarities of the observed samples and organize them into several clusters. Multidimensional scaling is the visualization of Cluster Analysis in a two dimensional coordinate system.

## 6. CASE STUDY

In this section, we illustrate how the statistical models described above can be used in modeling and developing operation policies for the UMTS network. Let's assume a network environment as shown in Figure 3. We use Figure 3 as a trial UMTS PS network. The system as highlighted is composed of a SGSN which connects with radio network via Iu-PS interface and a GGSN which accesses Internet and Intranet of Enterprise 1. Firewall and Network Address Translation (NAT) are built between UMTS PS network and external networks. The radio network domain consists of a Node-B (Base station) and a Radio Network controller (RNC). The network administrator monitors the network traffic through management station with authorities to access the network entities (NE) of UMTS PS network. The objective of this experiment is to monitor the throughput in interface Gi (Eth1:100) as it leaves GGSN.

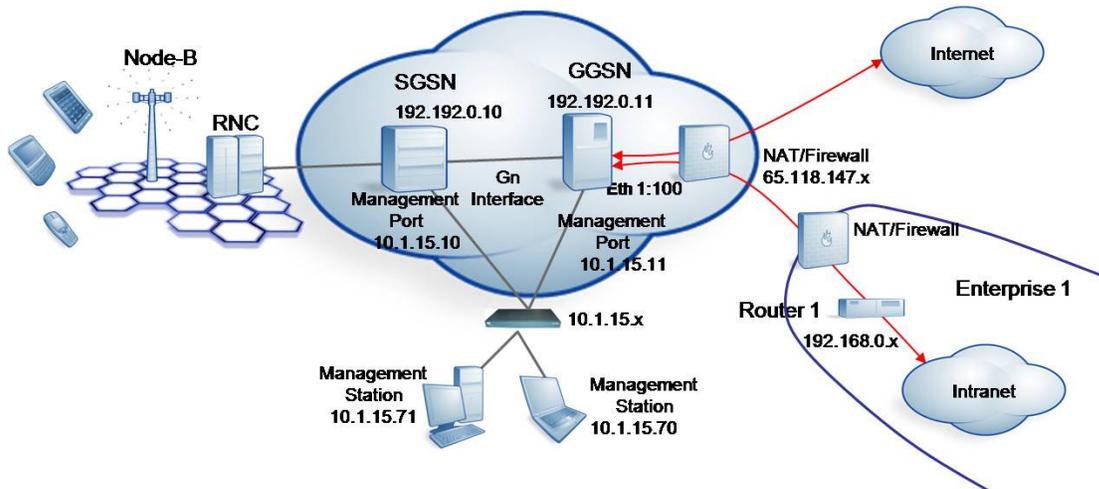

Figure 3. Network topology of trail UMTS PS network

The services are randomly triggered by a service/call generator tool in lieu of RNC, Node B and wireless terminals. The tool stores large quantities of historical traffic samples from a certain mobile operator A's network environment. Hence the tool in our case is actually a substitution of radio domain to simulate the real network environment of the mobile operator A. The simulated traffic generated by the tool is stochastically delivered into SGSN via Iu-PS interface and further transported through packet switched domain. The whole simulation process is no difference with a real network environment from traffic monitoring aspect. The performance parameters and service parameters are monitored as outputs based on the simulated traffic of services generated by the service generator. If this model is applied in a real environment, the monitored data will be the monitoring result based on the real traffic generated and delivered from radio domain. Five key performance indicators (KPI) are recorded as the network QoS parameters: Latency, GGSN average loading (Utilization), throughput in Eth1:100, packet loss in interface Gn (192.168.0.11) between GGSN and SGSN, and packet loss in interface Gi (Eth1:100 IP address: 192.168.0.12) between GGSN and external network. The management station collects the KPI data in 20 continuous sample periods (1 hour as 1 sample period). The sample data round up to fifth places of decimals after unit conversion from per hour to per second are recorded in Table 1 below.

Table 1. KPI data

| Sample period | GGSN utilization (%) | Gn interface Packet loss (Packet/s) | Gi interface Packet loss (Packet/s) | Latency (second) | Throughput in Gi interface(Eth1:100) (Mbps) |
|---|---|---|---|---|---|
| Hour 1 | 1 | 0 | 0 | 0.00204 | 2.442508 |
| Hr 2 | 1 | 0 | 0 | 0.00213 | 3.348526 |
| Hr 3 | 3 | 0.00028 | 0 | 0.00238 | 87.952500 |
| Hr 4 | 3 | 0 | 0 | 0.00243 | 99.157604 |
| Hr 5 | 5 | 0 | 0 | 0.00294 | 216.021441 |
| Hr 6 | 6 | 0 | 0.00028 | 0.00277 | 238.313785 |
| Hr 7 | 2 | 0 | 0 | 0.00208 | 28.812852 |
| Hr 8 | 2 | 0 | 0 | 0.00213 | 48.393216 |
| Hr 9 | 3 | 0.00333 | 0.00056 | 0.00217 | 65.983333 |
| Hr 10 | 2 | 0 | 0 | 0.00208 | 29.313644 |
| Hr 11 | 2 | 0 | 0.0025 | 0.00213 | 57.543637 |
| Hr 12 | 1 | 0 | 0 | 0.00200 | 2.781329 |
| Hr 13 | 1 | 0 | 0 | 0.00200 | 2.660693 |
| Hr 14 | 1 | 0 | 0 | 0.00200 | 2.667828 |
| Hr 15 | 1 | 0 | 0 | 0.00200 | 3.030091 |
| Hr 16 | 1 | 0 | 0 | 0.00200 | 2.578499 |
| Hr 17 | 1 | 0 | 0 | 0.00204 | 2.371938 |
| Hr 18 | 1 | 0 | 0 | 0.00213 | 2.370775 |
| Hr 19 | 1 | 0 | 0 | 0.00238 | 2.373311 |
| Hr 20 | 1 | 0 | 0 | 0.00243 | 2.369829 |

### 5.1. MULTIDIMENSIONAL SCALING, CLUSTERING AND CORRESPONDENCE ANALYSIS

Figure 3 displays the result of Cluster Analysis for mobile operator A's network: five clusters are formed via the complete linkage method which is a hierarchical clustering method. The visualization of the clusters can be displayed by Figure 4 via a Multidimensional Scaling Analysis. Since the clustering analysis forms five clusters from all 20 sample time spots, Figure 4 is similarly divided into 5 groups according to the result from Clustering Analysis.

As per the five clusters formed in the Cluster Analysis, the multidimensional scaling for the mobile operator A's UMTS PS trial network also presents five groups according to their similarities in the QoS KPIs. From Reference [11], we can calculate the cumulative proportion of the first two eigen values in our case is 82.15% and the elbow of the stress function occurs at 2, hence the fit into a two dimensional space is sufficient. In detail, the Figure 3 shows that the 20 sample hours forms five groups: Hour 9 as one group, Hour 11 as another group; Hour 3 and 4 as third group; Hour 5 and 6 as fourth group; and the rest sample hours for the fifth group.

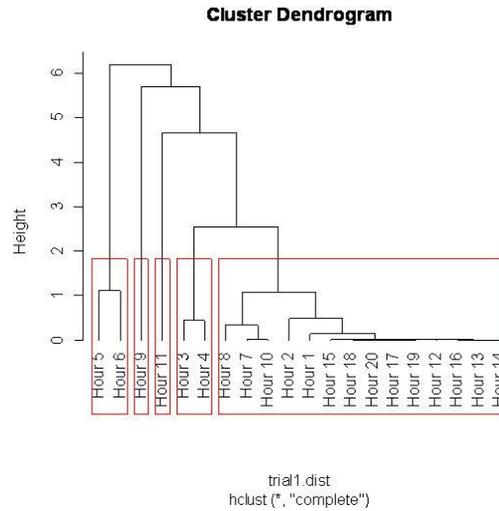

Figure 3. Cluster Analysis for Trial Network

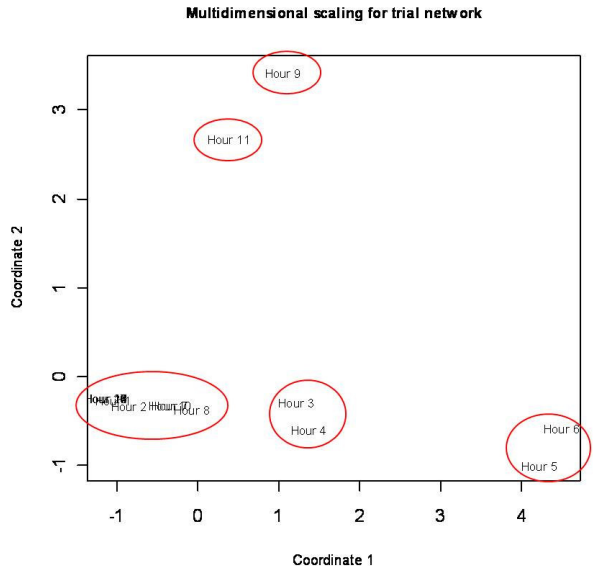

Figure 4. MDS for UMTS PS trial network

According to relative distances and similarities, the classification result presented from the Multidimensional Scaling is consistent with the result of Clustering Analysis, so the classification is reasonable. Furthermore, the common performance of the items (time spots) in a same group can be obtained based on the Correspondence Analysis for mobile operator A as illustrated in Figure 5.

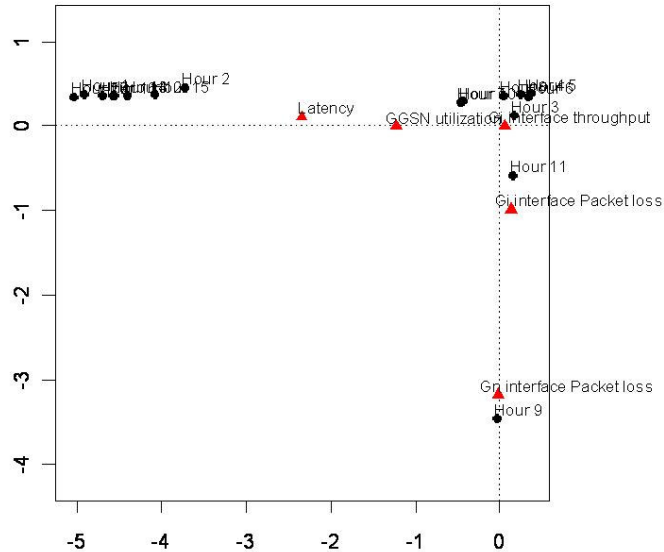

Figure 5. Correspondence Analysis for UMTS PS trial network

Now, let's examine the network performance.
*1) Packet Loss*

Figure 5 indicates that, in mobile operator A's network, Hour 9 is mainly suffering from Gn interface packet loss and Hour 11 is undergoing Gi interface packet loss. In general Hour 9 and 11 have same type of QoS issue: Packet loss, but existing in different interfaces. Associate Figure 5 with column 3 and 4 in Table 1, we obviously can find that during those two sample hours the UMTS PS packet loss ratio presents an increasing trend in which the packet loss ratio is increasing in Gn interface in hour 9 while that is increasing in Gi interface in hour 11. It may be noted that the column 3 and 4 of Table 1 contain many zeros since the monitoring data is rounded to the fifth place of decimal. So it doesn't denote the packet loss is zero in those corresponding sample hours but only presents the packet loss value is extremely low in those sample hours.

Packet loss can be caused by a number of factors such as physical links failure, faulty equipment, faulty network drivers, signal degradation, over-loaded links, corrupted packets denied in transport, or routing error, the first three of which are normally due to hardware failure and may result in the complete communication termination so it seldom occurs. If it occurs frequently in mobile operator A's network, it may indicate the links or equipment (SGSN and GGSN) has serious bugs in either hardware or software. For a long run operation, the mobile operators may consider rolling out new equipments or links to upgrade or replace the current ones.

The rest of the factors may be more familiar to network administrators. From Table 1 one can observe that in either hour 9 and 11 the communication is still continuous, hence the root cause of packet loss here should be network congestion or routing error. The policy to resolve this issue over a long run is to investigate which interfaces are undergoing heavy throughput: Once the interface is confirmed, then identify the types of fault through IP accounting on the problematic interface. If the issue is "attack", the solution is to block the attack via enabling access list. To resolve the other causes, such as excessive user datagram protocol (UDP) throughput, broadcast traffic, or DNS packets through the router, the solution is to implement throughput control to

International Journal of Next-Generation Networks (IJNGN), Vol.2, No.1, March 2010

reduce packet loss during the peak hours. However the packet loss resulted by this category of factors does not always indicate a problem. In UMTS PS, it sometime happens and causes highly noticeable performance downgrade or jitter with "heavy services" such as streaming services, voice over IP, online gaming and videoconferencing, and will affect all other network applications to a degree.

*2) Latency, Throughput and Utilization*

Hour 3, 4, 5 and 6 tend to be associated with throughput of Gi interface and GGSN utilization. It illustrates that the latency, loading, and utilization are relatively high in hour 3, 4, 5 and 6. The QoS decreases sharply in these four peak hours. Similarly, all the other hours tend to be associated with network latency and GGSN utilization.

Delay, jitter, utilization as well as packet loss are the primary quality measures in UMTS PS network. Based on these metrics, mobile operator A can add bandwidth to suffice increased traffic. But simply adding more bandwidth only for the limited period is not a wise choice. It suggests mobile operator A should reasonably allocate limited network resources to the clients with different priorities during the peak hours. GGSN as the network router implements QoS capabilities that prioritize data traffic across the UMTS PS network based on the types of service packets being routed. However, if the majority of traffic going through the GGSN is the same traffic such as streaming video it will all have high priority and network throughput and congestion will again be a problem. If the erosion of QoS (latency and jitter) is caused by contention, it may be an evidence that there are more packets than supportable bandwidth and extra packets are denied and blocked. A retransmission may burden the overloaded network again. In terms of this issue, a UMTS PS network expansion should be considered by mobile operator A to accommodate increased traffic.

### 5.2. CORRELATION ANALYSIS

In evaluating the performance of a UMTS PS network, we may monitor latency, throughput and packet loss which best reflect the real time QoS changes for individual services to the subscribers. In a UMTS PS network, data traffic can be categorized into four types: conversational traffic which includes voice or video call applications; streaming traffic which includes broadcast audio and video; interactive traffic includes web service, database or games; and background traffic includes E-mail and file transfers. Correlation analysis extracts the underlying relationship between traffic parameters and performance parameters.

Voice service is considered mainly going through circuit switched network but not through UMTS PS network. However voice over IP technology is available via UMTS PS in UMTS, so conversational traffic is still included in our case. In the trial, following services are considered: E-mail service, FTP service, web (http) service, audio streaming and video streaming services. The Pearson correlation analysis applies to investigate in our trial the impacts of service types on the QoS in UMTS PS network. The trial records five service traffic parameters with three performance (QoS) parameters in port eth1:100 per 60 seconds in 8 hours. Table 2 displays the 20 samples from the total 480 recorded samples.

Table 2. Sample data from the trial UMTS PS network

| Sample Period | Latency (Second) | Throughput (Mbps) | Packet Losses | Web service (Mbps) | Voice (Mbps) | FTP (Mbps) | E-mail | Video (Mbps) |
|---|---|---|---|---|---|---|---|---|
| Sample 1 | 1.219 | 26.89 | 1 | 3.376 | 1.12 | 5.779 | 0 | 6.170 |
| Sample 2 | 1.114 | 11.15 | 0 | 1.201 | 2.79 | 3.543 | 0.207 | 2.599 |
| Sample 3 | 1.115 | 11.16 | 0 | 1.101 | 1.59 | 2.450 | 0 | 2.413 |
| Sample 4 | 1.158 | 14.65 | 0 | 1.770 | 1.19 | 0 | 0 | 2.732 |
| Sample 5 | 1.155 | 14.65 | 0 | 0.792 | 3.56 | 2.769 | 0 | 1.910 |
| Sample 6 | 1.301 | 30.69 | 1 | 4.095 | 5.16 | 0 | 0.414 | 5.974 |
| Sample 7 | 1.293 | 30.69 | 1 | 6.010 | 3.31 | 5.981 | 0 | 6.316 |
| Sample 8 | 1.289 | 26.84 | 0 | 5.508 | 2.17 | 6.774 | 0.128 | 5.154 |
| Sample 9 | 1.281 | 26.84 | 1 | 7.965 | 1.72 | 5.853 | 0.327 | 5.780 |

| Sample 10 | 1.283 | 26.52 | 0  | 4.781 | 3.17 | 6.120  | 0     | 3.125  |
| Sample 11 | 1.285 | 26.52 | 0  | 3.790 | 6.03 | 7.002  | 0     | 4.538  |
| Sample 12 | 1.311 | 32.93 | 2  | 5.834 | 3.17 | 9.176  | 0     | 7.437  |
| Sample 13 | 1.309 | 32.93 | 1  | 7.860 | 6.58 | 8.402  | 1.379 | 11.508 |
| Sample 14 | 1.193 | 19.24 | 0  | 1.659 | 4.03 | 5.198  | 0.693 | 5.001  |
| Sample 15 | 1.201 | 19.24 | 0  | 1.878 | 5.24 | 4.708  | 0.145 | 4.609  |
| Sample 16 | 1.282 | 28.75 | 1  | 4.012 | 6.79 | 7.934  | 0     | 8.622  |
| Sample 17 | 1.291 | 28.75 | 1  | 4.665 | 6.69 | 0      | 0.019 | 7.480  |
| Sample 18 | 1.450 | 45.62 | 11 | 5.234 | 7.75 | 11.356 | 0.954 | 23.990 |
| Sample 19 | 1.499 | 45.62 | 7  | 6.651 | 6.51 | 16.435 | 0     | 24.105 |
| Sample 20 | 1.460 | 45.63 | 1  | 5.633 | 6.50 | 10.994 | 0     | 19.786 |

Table 3. Outputs of correlation analysis

| Latency | 1.0000000 | | | | | | | |
| Throughput | 0.9837184 | 1.0000000 | | | | | | |
| Packet Losses | 0.6901798 | 0.6891368 | 1.0000000 | | | | | |
| Web Service | 0.7485133 | 0.7612625 | 0.3635971 | 1.0000000 | | | | |
| Audio | 0.6506331 | 0.6358085 | 0.4987593 | 0.3042882 | 1.0000000 | | | |
| FTP | 0.7698489 | 0.7523677 | 0.6265178 | 0.5686828 | 0.4397341 | 1.0000000 | | |
| E-mail | 0.1698499 | 0.2191700 | 0.3366617 | 0.2943249 | 0.3575749 | 0.1519382 | 1.0000000 | |
| Video | 0.8914506 | 0.8857731 | 0.8393494 | 0.5322320 | 0.6656175 | 0.7937543 | 0.2991514 | 1.0000000 |
| | Latency | Throughput | Packet losses | Web services | Voice | FTP | E-mail | Video |

*1) Performance (QoS) Parameters-Positive Correlation*

Table 3 above shows the correlation between the variables. The three performance (QoS) parameters are highly correlated with each other: particularly the correlation between latency and throughput is close to 1. This indicates the network loading reflects a near perfect positive correlation to latency. So in UMTS PS network a heavy network loading normally results in a late response from the network entities to the service requested by the subscribers.

In mobile operator A's network, packet losses also present a close correlation (0.6901798) with latency and throughput, meaning that if there are too many users in UMTS PS network, congestion occurs and packets will be either dropped or require retransmission which takes time and impact overall QoS. It is also concluded that latency, throughput and packet losses can be categorized into the same type of QoS indicators in UMTS PS network. The correlation test shows the P-values between Latency and Throughput, between Latency and Packet losses and between Throughput and Packet losses are $7.105 \times 10^{-15}$, $7.573 \times 10^{-4}$, and $7.772 \times 10^{-4}$, all of which are statistically significant in 95% confident interval. Hence the correlation results are acceptable. Any two performance parameters are significantly correlated.

*2) Correlation between Service and Performance (QoS) parameters*

Table 3 also displays the relationship between service parameters and QoS parameters. Voice, FTP and Video service parameters, as the explanatory variable, are positively correlated with any of QoS parameters. The only exception is E-mail, which presents a low correlation to three QoS parameters. The reason is that E-mail service, unlike Audio, Video or FTP services, is a relatively time discrete service. The throughput generated by E-mail service is not as continuous as that by other streaming services such as Video services. Also the paroxysmal E-mail service does not pass the correlation test with any QoS parameters.

All the other streaming services are generally time continuous. So they are closely positive correlated to any of QoS parameters. Meanwhile the correlation test shows the statistical results are statistically significant at 95% confidence interval between any service parameters (other than E-mail) and any QoS parameters. It is concluded from Table 3 that time continuous services positively impact the QoS in the trial UMTS PS network. The time continuous services are the major factors that result in the network congestion, jitter and latency.

The result of correlation analysis reminds network administrators of monitoring and regulating the streaming services and http services before the network congestion and latency occur. Corresponding actions shall be implemented in advance such as limit http maximum connections, set a access list (ACL), prioritize the streaming services to subscribers, and set a upper bound in uplink or downlink for subscribers in particular peak time. For example, mobile operator must be able to deliver a high QoS to users paying for gold or platinum tiered services and limit such services as downloading and seeding bandwidth intensive P2P files.

Lastly the collection and correlation of these data into trends for analysis after the fact is valuable for future network planning purpose. Network administrators shall predict the possible time slots in which the network suffers from heavy loading from time discrete services such as E-mail service. Extra network resources shall be pre-allocated for these peak time slots.

### 5.3. DISCUSSION: HOW TO MAKE LONG RUN DECISION FROM MULTIVARIATE STATISTICAL ANALYSIS?

Figure 6 and 7 are two examples to capture the live throughput changes for a certain interface in the network. The green wave line denotes inbound throughput while the blue wave represents outbound throughput. The throughput plotted in two figures is the average value in 5 minutes and in 24 hours respectively. These two types of graph are common to the network administrators as visualized outputs for "short-run" (daily) and "long-run" (yearly).

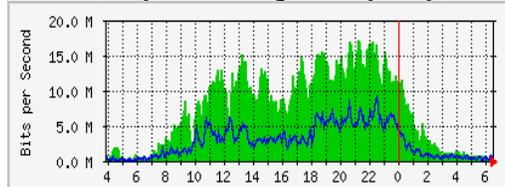

Figure 6. Throughput graph (average per 5 minutes)

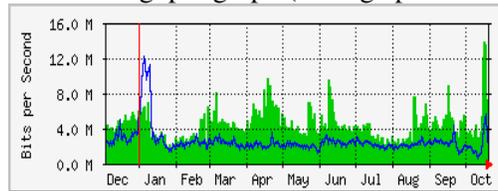

Figure 7. Throughput graph (average per 24 hours)

But why do we still need to extract KPIs to make further multivariate analysis for the long run? This can be explained by the fact that the graphs like Figures 6 and 7 can only visualize the quantitative changes of throughput whereas the multivariate statistical analysis for UMTS PS KPIs can determine the traffic parameters in the traffic model used for perspective network dimensioning.

In dimensioning the UMTs packet switched network, the mobile operators first need to form a traffic model to determine how much bandwidth shall be added based on the current throughput configuration of network entities in UMTS PS network or if any new network entities may be added into the existing network when the estimated bandwidth based on the traffic model exceeds the maximum configurable capability of current network entities. A traffic model with inexact estimated parameters may result in poor quality of service or even serious outage in the prospective network. Therefore, the accuracy of the traffic model actually decides the quality of service of the prospective network in a long run. The analysis in our paper will definitely improve the precision of the traffic parameters in the traffic model used for network dimensioning. It helps mobile operators obtain these need parameters in traffic model which are not visualized in a time-performance two dimensional coordinates system like Figure 6 and 7.

Creating a traffic model in Reference [19] is the premise of dimensioning a UMTS packet switched network. Reference [19] listed all the needed parameters in a traffic model for the capacity dimensioning of the UMTS PS network. The values of all those parameters can be estimated from our proposed multivariate analysis on the historical performance data. Reference [20] and [21] introduce the algorithms to dimension the UMTS packet switched and circuit switched networks. The parameters in the traffic model can be applied into the algorithms to generate the dimensioning results for the prospective UMTS PS network. Likewise, the multivariate analysis can also be applied to other mobile networks for the dimensioning work in a long run.

## 6. CONCLUSION

The paper first introduced the current monitoring method in "short run" and "long run" for UMTS PS network and analyzed the problem in current network O&M. Some proposed multivariate analysis methods are applied to UMTS PS KPIs to reveal some underlying results that current monitoring methods can not obtain. It helps network administrators effectively and actively forecast and prevent potential problems in packet switched network in advance. In addition, the analysis generates the needed parameters the use of the prospective network dimension. As a result, mobile operators are able to shape O&M policies for a long run.

In UMTS PS network operation and maintenance, instant network monitoring captures the real time parameters or indicators from UMTS PS network which enable network administrators to identify the network performance problems and adjust the system to address the problems in a short term.  However, the statistics based analysis extracts the performance characters of UMTS PS network for the long term by analyzing large quantities of network performance data that can help network administrators predict the potential trend of changes of network performance in advance. Preventive measures can be implemented to improve QoS of UMTS PS network; meanwhile the outputs from statistical analysis can also be used to support the network diagnosis. Mobile operators support a wide range of service types over UMTS PS network. Each type has a widely diverging traffic profile in terms of bit rate, maximum delay, maximum jitter, susceptibility buffer, and average packet size. The statistical performance analysis for those service profiles also generates the needed parameters of the traffic model for the mobile operators to plan and dimension their prospective UMTS PS network. Actually these multivariate analysis methods can be universally adopted in other types of networks. In summary, the proposed methods applied in the paper acquaint mobile operators with the long run network performance from a macro scope. Hence the macro analysis helps management make better strategic decisions in UMTS PS network evolution and transition to Evolved Packet Core (EPC).

## REFERENCE

[1] ITU-T I.363.5, *B-ISDN ATM Adaptation Layer Specification: Type 5 AAL - Series I: Integrated Services Digital Network Overall Network Aspects and Functions - Protocol Layer Requirements.*
[2] 3GPP TS23.060, General Packet Radio Service (GPRS); Service description; Stage 2
[3] 3GPP TS22.060, General Packet Radio Service (GPRS); Service description; Stage 1
[4] 3GPP TS24.008, Mobile radio interface layer 3 specification; Core Network Protocols – Stage 3
[5] 3GPP TS 23.002, Technical Specification Group Services and Systems Aspects; Network architecture.
[6] 3GPP TS 29.016, General Packet Radio Service (GPRS) ; Serving GPRS Support Node (SGSN) ̵Visitors Location Register (VLR); Gs interface network service specification.
[7] 3GPP TS29.018, General Packet Radio Service (GPRS); Serving GPRS Support Node (SGSN) ̵Visitors Location Register (VLR); Gs interface layer 3 specification.
[8] 3GPP TS29.060, General Packet Radio Service (GPRS); GPRS Tunnelling Protocol (GTP) across the Gn and Gp Interface.
[9] 3GPP TS29.061, General Packet Radio Service (GPRS); Interworking between the Public Land Mobile Network (PLMN) supporting GPRS and Packet Data Networks.
[10] 3GPP TS 32.015, Technical Specification Group Services and System Aspects; Telecommunication Management; Charging and billing; 3G call and event data for the Packet Switched domain.
[11] Richard A. Johnson, Dean W. Wichern, *Applied Multivariate Statistical Analysis*. Prentice Hall, 6th edition.
[12] T. W. Anderson (2003). *An Introduction to Multivariate Statistical Analysis*, Third Edition, Wiley.
[13] Abdelmonem A. Afifi, Virginia Clark, Susanne May (2004). *Computer-Aided Multivariate*

*Analysis*, Fourth Edition, CRC Press.

[14] Marcelo Resende, *Efficiency measurement and regulation in US telecommunications: A robustness analysis.* International Journal of Production Economics, Volume 114, Issue 1, July 2008, Pages 205-218

[15] Inmaculada Cava-Ferreruela, Antonio Alabau-Munoz, *Broadband policy assessment: A cross-national empirical analysis*, Telecommunications Policy, 2006 Vol.30 (No.8/9).

[16] Youssef, M., Abdallah, M., Ashok Agrawala, *Multivariate analysis for probabilistic WLAN location determination systems*, Mobile and Ubiquitous Systems: Networking and Services, 2005. MobiQuitous 2005. The Second Annual International Conference, 17-21 July 2005, Page(s): 353-362

[17] Shintaro Okazaki, *What do we know about mobile Internet adopters? A cluster analysis*, Information & Management, Volume 43, Issue 2, March 2006, Pages 127-141.

[18] Ouyang, Y. and Fallah, M.H. (2009) 'Evolving core networks from GSM to UMTS R4 version', *Int. J. Mobile Network Design and Innovation*, Vol. 3, No. 2, pp.93–102.

[19] Ouyang, Y. and Fallah, M.H., *A Study of Throughput for Iu-CS and Iu-PS Interface in UMTS Core Network*. Performance, Computing and Communications Conference, IEEE International. Dec 14-16, 2009.

[20] Ouyang, Y. and Fallah, M.H., *A Study of Throughput for Nb, Mc and Nc Interface in UMTS Core Network*. Performance, Computing and Communications Conference, IEEE International. Dec 14-16, 2009.

[21] Dong Hee Shin, *Overlay networks in the West and the East: a techno-economic analysis of mobile virtual network operators*. Telecommunication Systems, Volume 37, Number 4, 2008.

# APPENDIX

*A. Correlation*

The most preferred type of correlation coefficient is Pearson r, also called linear or product-moment correlation. Pearson correlation assumes that the two variables are measured on at least interval scales, and it determines the extent to which values of the two variables are "proportional" to each other. Proportional means linearly related; that is, the correlation is high if it can be "summarized" by a straight line. If the correlation coefficient is squared, then the resulting value ($r^2$, the coefficient of determination) denotes the proportion of the variability in response variable that is explained by the regression line. A larger $r^2$ value represents the response variables are more tightly coupled around the fitted regression line.

Assume explanatory variable $X=(x_1, x_2, x_3, …, x_m)'$ and response variable $Y=(y_1, y_2, y_3, …, y_n)'$. The minimal sum of error square ($Q_{min}$) is used to measure the similarities of two variables.

$$Q_{min} = \frac{\sum_{i=1}^{n}(y_i - \bar{y})^2}{n}\left\{1 - \frac{[\sum_{i=1}^{n}(x_i - \bar{x})(y_i - \bar{y})]^2}{\sum_{i=1}^{n}(x_i - \bar{x})^2 (y_i - \bar{y})^2}\right\} \quad (1)$$

in which we assume

$$r_{xy}^2 = \frac{[\sum_{i=1}^{n}(x_i - \bar{x})(y_i - \bar{y})]^2}{\sum_{i=1}^{n}(x_i - \bar{x})^2 (y_i - \bar{y})^2} \quad (2)$$

$$r_{xy} = \frac{\sum_{i=1}^{n}(x_i - \bar{x})(y_i - \bar{y})}{\sum_{i=1}^{n}(x_i - \bar{x})(y_i - \bar{y})} = \frac{cov(X,Y)}{sd(X)sd(Y)} \quad (3)$$

where $\bar{x}$ is the mean of X, $\bar{y}$ is the mean of Y, Cov denotes the covariance of X and Y, and Sd denotes the stand deviation of X.

$r_{xy}$=-1.0 denotes a perfect negative correlation while $r_{xy}$=+1.00 represents a perfect positive correlation. $r_{xy}$=0 means no linear relationship between X and Y.

*B. Factor analysis*

The purpose of factor analysis is to reduce the number of variables and to detect structure in the relationships between variables, which means to classify variables. Therefore, factor analysis is applied as a data dimension reduction or structure detection method. Factor analysis is to describe the covariance relationships among many variables in terms of a few underlying, but unobservable, random quantities (factors). Factor analysis can be considered an extension of principle component analysis, both attempting to approximate the covariance matrix Σ.

Assume observable vector X has mean $\mu$ and covariance $\Sigma$. We can obtain the factor analysis model in matrix notation below.

$$X = \mu + L * F + \varepsilon \quad (4)$$

in which $\mu_i$ represents the mean of variable i, $\varepsilon_i$ represents the ith specific factor, $F_j$ represents jth common factor, and $L_{ij}$ denotes the loading of the ith variable on the jth factor.

Assume $F_j$ and $\varepsilon_i$ are independent. Usually we have

$$E(F) = 0, \quad Cov(F) = I \quad (5)$$
$$E(\varepsilon) = 0, \quad Cov(\varepsilon) = \omega \quad (6)$$

Based on (4), we can obtain

$$\Sigma = Cov(X) = E(X - \mu)(X - \mu)' = LL' + \omega \quad (7)$$

There are several methods of estimation. Maximum likelihood method is introduced here since it's applied in the case of the paper.

If $F$ and $\varepsilon$ are assumed normally distributed, the maximum likelihood estimates of the factor loadings and specific variances may be obtained. The likelihood is represented by

$$L(\mu, \Sigma) = (2\pi)^{-np/2} |\Sigma|^{-n/2} e^{-\left(\frac{1}{2}\right) tr\left[\Sigma^{-1}\left(\sum_{j=1}^{n}(x_j - \bar{x})(x_j - \bar{x})' + n(\bar{x} - \mu)(\bar{x} - \mu)'\right)\right]} \quad (8)$$

To make L well defined, impose the computationally convenient uniqueness condition

$$L'\omega^{-1}L = \Delta \quad (9)$$

In which $\Delta$ is a diagonal matrix. To maximize the L in formula (8), we can obtain the maximum likelihood estimates $\hat{L}$ and $\hat{\omega}$.

*C. Multidimensional Scaling*
Please refer to chapter 12 in Reference [11].

*D. Correspondence analysis*
Please refer to chapter 12 in Reference [11].

*E. Cluster Analysis and Distance Measures*
   Cluster analysis is an exploratory data analysis tool which aims at sorting different objects into groups in a way that the degree of association between two objects is maximal if they belong to the same group and minimal otherwise. Cluster analysis simply discovers structures in data without explaining why they exist.
   The joining or tree clustering method uses the dissimilarities (similarities) or distances between objects when forming the clusters. Similarities are a set of rules that serve as criteria for grouping or separating items. These distances (similarities) can be based on a single dimension or multiple dimensions, with each dimension representing a rule or condition for grouping objects. The most straightforward way of computing distances between objects in a multi-dimensional space is to compute Euclidean distances. If we had a two- or three-dimensional space this measure is the actual geometric distance between objects in the space.

*Euclidean Distance*
   This is probably the most commonly chosen type of distance. It simply is the geometric distance in the multidimensional space. It is computed as Distance(x,y) = $\{\sum_i (x_i - y_i)^2\}^{1/2}$
   Note that Euclidean (and squared Euclidean) distances are usually computed from raw data, and not from standardized data. This method has certain advantages (e.g., the distance between any two objects is not affected by the addition of new objects to the analysis, which may be outliers). However, the distances can be greatly affected by differences in scale among the dimensions from which the distances are computed.

*Squared Euclidean Distance*
   Square the standard Euclidean distance in order to place progressively greater weight on objects that are further apart. This distance is computed as Distance(x,y) = $\sum_i (x_i - y_i)^2$

*City-block (Manhattan) Distance*

This distance is simply the average difference across dimensions. In most cases, this distance measure yields results similar to the simple Euclidean distance. However, note that in this measure, the effect of single large differences (outliers) is dampened (since they are not squared). The city-block distance is computed as Distance(x,y) = $\sum_i |x_i - y_i|$

*Chebychev Distance*

This distance measure may be appropriate in cases when one wants to define two objects as "different" if they are different on any one of the dimensions. The Chebychev distance is computed as Distance(x,y) = Max $|x_i - y_i|$

*Power Distance*

Increase or decrease the progressive weight that is placed on dimensions on which the respective objects are very different. This can be accomplished via the power distance. The power distance is computed as Distance(x,y) = $\sum_i (|x_i - y_i|^p)^{1/r}$ where r and p are user-defined parameters. A few example calculations may demonstrate how this measure "behaves." Parameter p controls the progressive weight that is placed on differences on individual dimension, parameter r controls the progressive weight that is placed on larger differences between objects. If r and p are equal to 2, then this distance is equal to the Euclidean distance.